\documentclass[aps,prb,twocolumn,showpacs,amsmath]{revtex4}
\usepackage[colorlinks=true,citecolor=blue]{hyperref}
\usepackage{graphics} 
\usepackage{epsfig}
\begin{document} 
   \title{
Heat production and current noise for single- and double-cavity quantum capacitors
         }
\author{
Michael Moskalets$^{1,2}$, 
and Markus B\"uttiker$^{1}$
}
\affiliation{
$^1$D\'epartement de Physique Th\'eorique, Universit\'e de Gen\`eve, CH-1211 Gen\`eve 4, Switzerland \\
$^2$Department of Metal and Semic. Physics, NTU "Kharkiv Polytechnic Institute", 61002 Kharkiv, Ukraine
}
\date\today
   \begin{abstract}
We analyze the frequency-dependent noise and the heat production rate for a dynamical quantum capacitor in the regime in which it emits single particles, electrons and holes. At low temperature and slow driving the relaxation resistance quantum, $R_{q} = h/(2e^2)$, defines the heat production rate in both the linear and non-linear response regimes. If a double-cavity capacitor emits particles in pairs, the noise is enhanced. In contrast the energy dissipated is suppressed or enhanced depending on whether an electron-hole pair or an electron-electron (a hole-hole) pair is emitted.
   \end{abstract}
\maketitle

\begin{figure}[b]
  \vspace{0mm}
  \centerline{
   \epsfxsize 8cm
   \epsffile{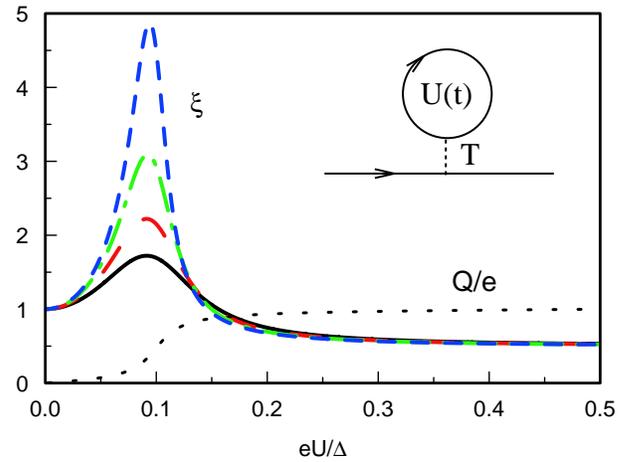}
             }
  \vspace{0mm}
  \nopagebreak
\caption{(color online) Inset: The cavity with level spacing $\Delta$, driven by the potential $U(t) = U \cos(\Omega t)$ is coupled to a linear edge state by the QPC with transmission $T$. Main: The normalized noise to dissipation ratio $\xi$ at zero temperature is shown as a function of the amplitude $U$.The quantity $\xi \equiv \xi(U)$ is defined as, $U^2 {\cal P}_{0}(\omega) /I_{E} = \xi(U) (2\hbar\omega^3/\Omega^2)\coth\left( \hbar\omega/2k_B\theta \right)$. The curves differ in transmission of the QPC: $T = 0.5$ (black solid), $0.4$ (red dashed), $0.3$ (green dot-dashed), $0.2$ (blue short-dashed). The charge $Q$ (dotted line) emitted during half of a period indicates the quantized emission regime at $eU \gtrsim 0.2\Delta$. The linear response regime is at $U \to 0$. }
\label{fig1}
\end{figure}

\begin{figure}[b]
  \vspace{0mm}
  \centerline{
   \epsfxsize 8cm
   \epsffile{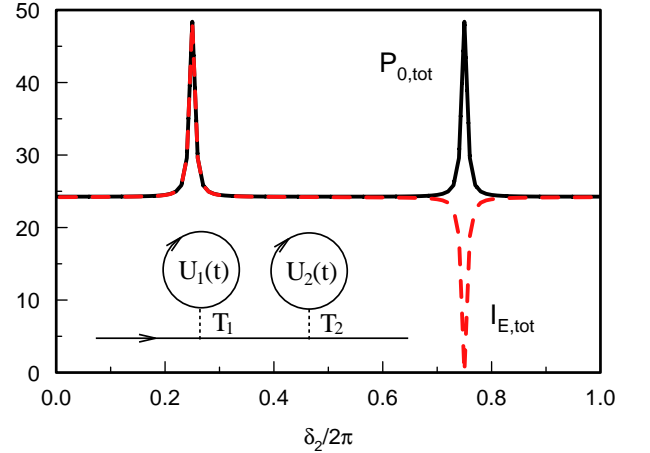}
             }
  \vspace{0mm}
  \nopagebreak
\caption{(color online) Inset: Two cavities with level spacing $\Delta_j$ ($j = 1,2$) are connected to a common edge state by QPCs with transmission $T_{j}$. Main: The heat current $I_{E,tot}$ in units of $\pi (eU_1 /\Delta_1 )^2 \hbar\Omega^2$ and the normalized noise power $P_{0,tot}$, both at zero temperature, are shown in function of the phase 
$\delta_{2}$ of the potential $U_{2}(t)$ 
in the quantized emission regime. The quantity $P_{0,tot}$ is defined as follows, ${\cal P}_{0,tot}(\omega) = P_{0,tot} (\pi e^2 \hbar^2 \omega^3 /\Delta_1^2)\coth\left( \hbar\omega/2k_B\theta \right) $. The parameters are: $T_j = 0.1$; $eU_j = 0.5\Delta_1$; $\Delta_2 = \Delta_1$, $\delta_{1} = \pi/2$.
}
\label{fig2}
\end{figure}

\textit{Introduction.}-- Recent experiments demonstrate \cite{Gabelli06,Feve07} that a quantum capacitor \cite{BTP93} in a two-dimensional electron gas in the integer quantum Hall effect regime is a promising device for the realization of a sub-nanosecond, few-electron, coherent quantum electronics. 
The capacitor, shown in the inset of Fig.\,\ref{fig1}, serves as an RC circuit with a quantized charge-relaxation resistance \cite{BTP93,Gabelli06,NLB07}.
This quantization suggests a high-frequency charge detector with near quantum limited efficiency. \cite{detectorNB09}
The quantum capacitor can also be used as a single-particle emitter. \cite{Feve07} 
Using capacitors as emitters, several effects were predicted including shot noise plateaus \cite{OSMB08}, particle emission and reabsorption \cite{SOMB08}, and a tunable two-particle Aharonov-Bohm effect \cite{SMB09}.  
With increasing frequency the quantum capacitor can exhibit an inductive-like response. 
\cite{indWWG07,ZJY08} 
\\ \indent
One of the questions important for any electronic device is how noisy it is and how much energy is dissipated while it is working. 
Answering these questions we also get more insight into relevant physical processes. 
Our aim is to explore the electrical noise and the energy loss of the capacitor driven by a periodical potential. 
Note that the capacitor by itself can not produce a zero-frequency noise \cite{BB00}, therefore, the noise, we are interested in, is a frequency-dependent one. 
In contrast, the dissipated energy can be characterized with the help of a rate $I_{E}$, i.e, the energy flow averaged over the period of a drive ${\cal T}$.   
In the linear response regime, in accordance with the standard fluctuation-dissipation theorem \cite{CW51}, both the frequency-dependent noise and the heat production rate are governed by the same quantity, the real part of an admittance, which, for a single-channel capacitor at zero temperature and for slow driving, is a universal quantity, $R_{q} = h/(2e^2)$, independent of parameters of the capacitor, as it was predicted theoretically \cite{BTP93} and revealed experimentally \cite{Gabelli06}. 
However, for the non-linear response regime in which a single-particle emission can be achieved \cite{Feve07}, such a simple relation is not applicable. 
This regime is of our prime interest here.
\\ \indent
The response of a cavity, driven by a potential $U(t)$ with large frequency $\Omega$, to an additional small amplitude excitation with smaller frequency was addressed in Ref.\,\onlinecite{ParkAhn08}. 
In contrast, we are interested in both the noise measured at frequency $\omega$ and the steady heat production rate solely due to the potential $U(t)$.
\\ \indent
\textit{The model and the quantities of interest.}-- 
We consider a capacitor consisting of a single cavity (see inset of Fig.\,\ref{fig1}) or several cavities (see inset of Fig.\,\ref{fig2}) placed in series and coupled to the same edge state. 
The cavity consists of a circular edge state coupled via a quantum point contact (QPC) with reflection/transmission probability ${\mathrm r}/\tilde {\mathrm t}$ to an edge state which in turn is connected to a metallic reservoir with equilibrium electrons described by the Fermi distribution function $f_{0}(E)$ with chemical potential $\mu$ and temperature $k_B\theta$. \cite{Gabelli06,Feve07,MSB08,KSL08}
The potential varying in time with period ${\cal T} = 2\pi/\Omega$ changes the position of quantum levels in the cavity vis a vis the Fermi level.  
Scattering of electrons propagating in the linear edge state past a periodically driven capacitor is described by the Floquet amplitude $S_{F}(E_{m}, E)$ for a carrier incident with energy $E$ which absorbs an energy $m\hbar\Omega = E_{m} - E$. 
It is convenient to write this amplitude as the Fourier transformation, $S_{F}(E+m\hbar\Omega,\, E) = \int_{0}^{\cal T} (dt/{\cal T}) e^{im\Omega t}S_{in}(t,E)$. 
Then the symmetrized current correlation function \cite{BB00} $P(\omega,\omega^{\prime})$ can be written as follows \cite{MB07}, 
${ P}(\omega,\omega^{\prime}) = \sum_{l=-\infty}^{\infty} 2\pi \delta(\omega + \omega^{\prime} - l\Omega) {\cal P}_{l}(\omega)$ with 
\begin{eqnarray}
{\cal P}_{l}(\omega) &=& \frac{e^2}{2h} \sum\limits_{n=-\infty}^{\infty} \int dE F\left(E,E_{n} - \hbar\omega\right)  \nonumber \\
\label{01}  
&&\times {\Pi}_{n}(E_{n} - \hbar\omega, E) {\Pi}_{l-n}(E,E_{n} - \hbar \omega)\,,
\end{eqnarray}
where
$F(E,E^{\prime}) = f_{0}(E)[1-f_{0}(E^{\prime})] + f_{0}(E^{\prime})[1-f_{0}(E)]$. and
${\Pi}_{q}(E^{\prime},E) = \left\{S_{in}^{\star}\left(t,E^{\prime}\right) S_{in}(t,E) -1 \right\}_{q}$. 
The lower index $_{q}$ denotes the Fourier coefficient.
The noise power possesses the following symmetry properties, ${\cal P}_{l}( \omega) ={\cal P}_{l}(l\Omega - \omega) $ and ${\cal P}_{l}(\omega) = {\cal P}_{-l}(-\omega)$. 
\\ \indent
The corresponding equations for the heat flow $I_{E}$ and the time-dependent charge current $I(t)$ are, 
\begin{equation}
I(t) = \frac{e}{h} \int dE f_{0}(E) \left\{ \left | S_{in}(t,E) \right |^2 - 1  \right\} , \label{02} 
\end{equation}
\begin{equation}
I_{E} = - \frac{i}{2\pi} \int dE\,f_{0}(E) \int_{0}^{\cal T} \frac{dt}{\cal T}\,S_{in}(t,E)\, \frac{\partial S_{in}^{\star}(t,E) }{\partial t} , \label{03} 
\end{equation}
\noindent
All equations (\ref{01}) - (\ref{03}) are valid for arbitrary frequency $\Omega$ and arbitrary amplitude of the potential $U(t)$.
The amplitude $S_{in}$ valid for $\hbar\Omega \ll \mu$ for a single- and double-cavity capacitors is given in Refs.\,\onlinecite{MSB08} and \onlinecite{SOMB08}, respectively.
\\ \indent
In what follows we are interested in the slow frequency (adiabatic) regime when $S_{in}(t,E)$ is expressed in terms of the frozen scattering amplitude $S(t,E)$, see Ref.\,\onlinecite{MB05},
\begin{equation}
S_{in}(t,E) \simeq S(t,E) + \dfrac{i\hbar}{2} \dfrac{\partial^2 S(t,E) }{\partial t \partial E } + \hbar\Omega A(t,E)\,,
\label{04}
\end{equation}
\noindent
with anomalous amplitude $A(t,E)$ satisfying
\begin{equation}
2\hbar\Omega\, {\mathrm Re} \left\{ S^{*} A \right\} = \dfrac{i\hbar}{2} \left\{ \dfrac{\partial S^{*} }{\partial t } \dfrac{\partial S^{} }{\partial E } - \dfrac{\partial S^{*} }{\partial E } \dfrac{\partial S^{} }{\partial t } \right\}.
\label{05}
\end{equation}
\noindent
The frozen amplitude is a stationary $S$-matrix element calculated for fixed $U$, with subsequent substitution $U \to U(t)$.
For the single-channel capacitor a unitary $S$ requires $|S(t,E)|^2 = 1$, hence, $S(t,E) = \exp\{i\varphi(t,E)\}$.
From Eqs.\,(\ref{04}), (\ref{05}), it follows that the non-adiabatic corrections are small, if the quantum $\hbar\Omega$ is smaller than some energy $\delta E$ characteristic for the stationary scattering amplitude, $\hbar\Omega \ll \delta E$. \cite{MB07} 
For the frequency-dependent noise, the adiabatic regime implies similarly $\hbar|\omega| \ll \delta E$. 
\\ \indent
In the lowest order in $\Omega$ the heat flow and the current are given in terms of the frozen scattering amplitude, \cite{MB07}
\begin{eqnarray}
I_{E} &=& \dfrac{\hbar}{4\pi} \int\limits dE \left( - \dfrac{\partial f_{0}}{\partial E} \right) \int\limits_{0}^{\cal T} \dfrac{dt}{\cal T} \left|\dfrac{\partial S(t,E)}{\partial t} \right|^2 , \label{06}  \\
I(t) &=& -\frac{i e}{2\pi} \int\limits dE \left( - \dfrac{\partial f_{0}}{\partial E} \right) S(t,E)\, \frac{\partial S^{\star}(t,E) }{\partial t}\,. \label{07} 
\end{eqnarray}
\noindent
At low temperatures the heat flow is related to the electric current, i.e., the measurement of $I_{E}$ can be done via the measurement of $I(t)$.
To show it we consider the average \cite{SOMB08},
$\left\langle I^2 \right\rangle = \int_{0}^{\cal T} (dt/{\cal T}) I^2(t)$. 
At $k_B\theta \ll \delta E$ in Eqs.\,(\ref{06}),  (\ref{07}) we neglect the energy dependence of the frozen amplitude and calculate it at the Fermi energy $\mu$. 
To simplify $\left\langle I^2 \right\rangle$ we use, $S \partial S^*/\partial t = - S^* \partial S/\partial t$, following from	the unitarity condition and finally obtain, 
\begin{equation}
I_E = R_{q} \left\langle I^2 \right\rangle .
\label{08}
\end{equation}
\noindent
Thus in the adiabatic low-temperature regime the heat produced by the dynamical capacitor is nothing but the Joule heat due to the relaxation resistance quantum \cite{BTP93,Gabelli06} $R_{q} = h/(2e^2)$. 
We are not expecting such a relation at finite temperatures and/or out of the adiabatic regime.\\ \indent
The noise power calculations are different for the single-cavity and the double-cavity cases.
\\ \indent
\textit{The single-cavity capacitor.}-- If the capacitor comprises only one cavity with a circular edge state driven by the uniform potential $U(t) = U\cos\left( \Omega t \right)$, then $\varphi(t,E) \equiv  \varphi(U(t),E)$.
In this case the right hand side of Eq.\,(\ref{05}) is zero and we conclude that $A = 0$. 
\\ \indent
To relate ${\cal P}_{l}(\omega)$ to $I_{E}$ we express both of them  in terms of the frozen density of states (DOS),
\begin{eqnarray}
\nu(t,E) &=& (i/2\pi) S(t,E) \partial S^{*}(t,E)/\partial E\,.
\label{09}
\end{eqnarray}
\noindent
Using Eq.\,(\ref{04}) with $A=0$ into Eq.(\ref{01}) and expanding $S(t,E_{n} - \hbar \omega) \approx S(t,E) + \hbar [n\Omega-\omega]\, \partial S/\partial E$ we find, 
\begin{equation}
{\cal P}_{l}(\omega) = \xi_{l}(\omega) \sum_{n=-\infty}^{\infty}
\int dE  F\left(E,E_{n} - \hbar \omega\right)  \nu_{n} \nu_{l-n}, 
\label{10} 
\end{equation}
\noindent
where $\xi_{l}(\omega) = (he^2/2) \omega (\omega-l\Omega)$, and $\nu_{q}$ is a Fourier coefficient for the frozen DOS.
\\ \indent
If
$k_B\theta \gg \hbar|\omega|, \hbar\Omega$ we use $F(E, E_{n} - \hbar \omega) \simeq -2 k_B\theta f_{0}^{\prime}$, where $f_{0}^{\prime} \equiv \partial f_{0}/\partial E$, and if $\hbar|\omega| \gg k_B\theta, \hbar\Omega$ we use $F(E, E_{n} - \hbar \omega) \simeq \left\{ f_{0}(E) - f_{0}(E_{n} - \hbar\omega) \right\}^2$.
Integrating over energy in Eq.\,(\ref{10}) we find for $k_{B}\theta \gg \hbar\Omega$ or $|\omega| \gg \Omega$:
\begin{equation}
{\cal P}_{l}(\omega) = \xi_{l}(\omega)
\hbar\omega \coth\left( \frac{\hbar\omega}{2k_{B}\theta} \right) \int dE \left( - f_{0}^{\prime} \right) \{\nu^{2}\}_{l}. 
\label{11} 
\end{equation}
\indent
If the single uniform potential with amplitude $|eU| \ll \mu$ acts onto the capacitor, we use, $\partial S/\partial t = - (\partial S/\partial E) edU(t)/dt$, 
to calculate the heat flow $I_{E}$. 
From Eq.\,(\ref{09}) we find, $\left| \partial S/\partial t \right|^2 = 4\pi^2 e^2 (dU/dt)^2 \nu^2(t,E)$.
Using the latter relation in Eq.\,(\ref{06}) we find for the harmonic potential $U(t) = U \cos\left( \Omega t \right)$,
\begin{equation}
\hspace*{-5pt} I_{E} =  C \int dE\,\left( - f_{0}^{\prime} \right)  \left( 2 \left\{ \nu^{2} \right\}_{0} - \left\{ \nu^{2} \right\}_{2} - \left\{ \nu^{2} \right\}_{-2} \right),
\label{12}
\end{equation}
\noindent
where $C = he^2 U^2 \Omega^{2}/8$.
Comparing Eq.\,(\ref{12}) with Eq.\,(\ref{11}) we find the following relation, 
\begin{eqnarray}
\hspace*{-2pt}{\cal P}_{0}(\omega) - \frac{ {\cal P}_{2}(\omega) + {\cal P}_{-2}(\omega) }{2} = \dfrac{2 I_{E} }{U^2 } \frac{\hbar \omega^{3} }{\Omega^{2} }  \coth \left( \dfrac{\hbar\omega}{2k_{B}\theta} \right)\hspace*{-3pt}. \label{13} 
\end{eqnarray}
\noindent
This relation is independent of the parameters characterizing a capacitor. 
It extends the fluctuation - dissipation theorem \cite{CW51} for the linear response regime to the non-linear regime of a single-channel capacitor and to measurement frequencies different from the driving frequency. 
\\ \indent
\textit{Quantized emission regime.}-- If the amplitude $U$ is comparable with the level spacing $\Delta$ for electrons in the cavity, then the regime of quantized emission can be achieved. \cite{Feve07} 
In this regime one (or several) electron(s) and hole(s) are emitted. \cite{Feve07,MSB08}
We assume the transmission of the QPC connecting the cavity to the linear edge state is small, 
$T\equiv |\tilde {\mathrm t}|^2 \ll 1$, 
such that the emission of an electron and a hole is 
separated in time.
Also we assume the temperature is smaller than the inverse of the half-width of an emitted current pulse, $k_{B}\theta\ll \hbar/\Gamma$ to neglect the temperature averaging. 
We choose the amplitude $U$ and the position of some energy level in the cavity $\epsilon_{k} = \mu + eU_{0}$ such that only this level crosses the Fermi level $\mu$ during the period. 
Then one electron and one hole are emitted at times $t_{\mp} = \mp t_{0}$.
The emission times are defined as follows,
$\epsilon_{k} + eU\left(t_{\mp}\right) = \mu$, 
with $\Omega t_{0} = \arccos\left(- U_{0}/U \right)$.
The corresponding DOS reads for $0 < t \leq {\cal T}$:
\begin{equation}
\nu(t,\mu) = 4\Gamma^2(\Delta T)^{-1} \sum_{\alpha = -,+} \left\{\left(t - t_{\alpha} \right)^{2} + \Gamma^{2} \right\}^{-1}  , \label{14}
\end{equation}
\noindent 
where $\Omega\Gamma =T\Delta/(4\pi |e| \sqrt{U^{2} - U_{0}^{2} })$. 
The emitted electron and hole are separated in time if $t_{0} \gg \Gamma$.
In this regime from Eqs.\,(\ref{12}) and (\ref{14}) we calculate the heat flow,
\begin{equation}
{\cal T} I_{E} = \hbar \Gamma^{-1}\,.
\label{15}
\end{equation}
\noindent
This heat flow is due to energy $\hbar/(2\Gamma)$ carried by both electrons or holes emitted during the period ${\cal T} = 2\pi/\Omega$.
\\ \indent
To calculate ${\cal P}_{l}(\omega)$, Eq.\,(\ref{11}), we need the Fourier coefficients for the squared DOS, Eq.\,(\ref{14}). We find,  
$\left\{\nu^{2} \right\}_{n} = (4\Omega\Gamma/\Delta^{2} T^{2}) e^{-|n|\Omega\Gamma}\ \left\{ e^{in\Omega t_{-}} +\, e^{in\Omega t_{+}}\right\}$, in leading order in $\Omega\Gamma \ll 1$.
Using it in Eq.\,(\ref{11}) we calculate the noise power for $\hbar/\Gamma\gg k_{B}\theta \gg \hbar\Omega$ or $\hbar/\Gamma \gg \hbar|\omega| \gg \hbar\Omega$:
\begin{equation}
\hspace*{-5pt} {\cal P}_{l}(\omega) = {\bar {\cal P}}_{0}(\omega) \left( 1 - l\Omega/\omega \right) (\Omega \Gamma/2) e^{-|l|\Omega\Gamma}  \cos\left(l\Omega t_{0}\right), 
\label{17} 
\end{equation}
\noindent
where ${\bar {\cal P}}_{0}(\omega) = 16 \pi e^2 \hbar^2 \omega^3 /(\Delta T)^2 \coth \left( \hbar \omega / 2 k_B \theta \right)$ is the maximum noise produced by a stationary capacitor. 
In the stationary case the noise is maximum if one of the capacitor's levels  aligns with the Fermi energy, $\epsilon_{k} = \mu$.  
Note that Eqs. (\ref{15}) and (\ref{17}) do satisfy Eq.\,(\ref{13}).
\\ \indent
It is instructive to compare the linear response and the quantized emission regimes. 
We choose $\epsilon_{k} = \mu$ to get the maximum noise in the former case, which is now realized if $\kappa = |eU|/\delta E \ll 1$, where $\delta E = \Delta T/(2\pi)$ is a level width. In this case $I_{E,lin} = (2/\pi) \hbar \Omega^2 \kappa^2$ and ${\cal P}_{0,lin}(\omega) = {\bar {\cal P}}_{0}( \omega)$.
In the quantized emission regime we calculate from Eqs.\,(\ref{15}), (\ref{17}), $I_{E,quan} = \hbar \Omega^2 \kappa / \pi$ and ${\cal P}_{0,quan} = {\bar {\cal P}}_{0}( \omega) /(4 \kappa)$.
Comparing these results we find,
\begin{equation}
\hspace*{-5pt} \left\{ U^2 {\cal P}_{0}(\omega)/I_{E}  \right\}_{(quan)} = (1/2) \left\{ U^2 {\cal P}_{0}(\omega) /I_{E}  \right\}_{(lin)}\,.
\label{18}
\end{equation}
\noindent
In the quantized emission regime, the noise to dissipation ratio is suppressed compared to the linear response regime, when the electron-hole pairs are emitted rather than separate particles, Fig.\,\ref{fig1}.
This ratio is enhanced in the transition region between plateaus, as it is shown in Fig.\,\ref{fig1} for the transition to the first plateau. 
In this case an electron and a hole are emitted nearly simultaneously. 
That suppresses their contribution to both the charge current and the heat current $I_{E}$, hence it increases the ratio $U^2 {\cal P}_{0}(\omega)/I_{E}$. 
The suppression of $I_{E}$ due to electron-hole annihilation will be clarified in the next section.
\\ \indent
\textit{The double-cavity capacitor.}-- 
The scattering amplitude $S_{in,tot}(t,E)$ for a capacitor comprising two cavities connected in series was introduced in Ref.\,\onlinecite{SOMB08}. 
Each cavity is driven by the corresponding potential $U_{j}(t) = U_{j} \cos\left( \Omega t + \delta_{j} \right)$, $j = 1,2$.
The general Eqs. (\ref{01}) -- (\ref{03}) and Eqs.(\ref{06})--(\ref{08}) in the adiabatic case, 
remain valid for a double-cavity capacitor.
In the adiabatic regime the heat flow in terms of the DOS $\nu_{j}$ of the cavities reads,
\begin{equation}
I_{E,tot}   =   \frac{he^2}{2} \int dE \left(- f_{0}^{\prime}\right) \int_{0}^{\cal T} \frac{dt}{\cal T} \left( \sum_{j=1}^{2} \nu_{j} \frac{dU_{j} }{dt} \right)^2. 
\label{19} 
\end{equation}
\indent
In the quantized emission regime when each cavity emits one electron and one hole during  a period $\cal T$, we use $\left\langle I^2 \right\rangle$ from Ref.\,\onlinecite{SOMB08}, and then from Eq.\,(\ref{08}) we find,
\begin{equation}
\begin{array}{l}
I_{E,tot} = \frac{\Omega}{4\pi} \left( \frac{\hbar}{\Gamma_{1}} + \frac{\hbar}{\Gamma_{2}} \right) \left\{2 - L\left(\Delta t_{1,2}^{(-+)} \right) \right.  \\
\ \\
\left. - L\left(\Delta t_{1,2}^{(+-)} \right) + L\left(\Delta t_{1,2}^{(--)} \right) + L\left(\Delta t_{1,2}^{(++)} \right) \right\},
\end{array}
\label{20} 
\end{equation}
\noindent
where $\Gamma_{j}$ is a half-width of a current pulse emitted by the $jth$ cavity, $\Delta t_{1,2}^{(\alpha \beta)} = t_{1\alpha} - t_{2\beta}$ is the difference of emission times, and $L(X) = 4\Gamma_{1}\Gamma_{2}/\left( X^2 + \left(\Gamma_{1} + \Gamma_{2}  \right)^2 \right)$.
\\ \indent
Remarkably, if the capacitor acts as a two-electron (two-hole) emitter, $\Delta t_{1,2}^{(\alpha\alpha)} = 0$, then the dissipated heat is enhanced compared to the regime when particles are emitted at different times. 
In contrast, if the capacitor emits electron-hole pairs, $t_{1,2}^{(-+)} = 0$ and/or $t_{1,2}^{(+-)} = 0$, then the generated heat is suppressed. 
The enhancement of $I_{E,tot}$ can be understood as an additional work generated by the external potentials $U_{j}(t)$ to inject two electrons (holes) above (below) the Fermi see into the same edge state. 
Therefore, the emitted pair of electrons (holes) has an energy larger than two particles emitted separately.   
The electron-hole pair emission can be viewed as an reabsorption by the second cavity of a particle emitted by the first cavity. Therefore, none of the particles carry energy out of the capacitor, hence $I_{E,tot}$ is suppressed.
Within this picture we can also say, that the power done by the first potential, $U_{1}(t)$, to inject a particle into the edge state was transferred and used in the second cavity to work against the second potential, $U_{2}(t)$. 
This is a realization of the general idea of {\it work transfer} in a coherent electron system put forward in Ref.\,\onlinecite{lili08}. 
\\ \indent
In contrast to the heat flow, the noise does not vanish even in the electron-hole emission regime. 
To calculate the noise power we need to obtain a corresponding adiabatic expansion, by analogy with Eq.\,(\ref{04}), for the scattering amplitude $S_{in,tot}(t,E)$. 
The frozen scattering amplitude $S_{tot}(t,E)$ is a product of frozen amplitudes for cavities, $S_{j}(t,E) = \exp\left\{i \varphi (U_j(t), E)  \right\}$, such that $S_{tot}(t,E) = S_{1}(t,E) S_{2}(t,E)$. 
For simplicity we neglect the contribution due to the wire connecting the cavities.
Expanding $S_{in,tot}(t,E)$ from Ref.\,\onlinecite{SOMB08} to linear in $\Omega$ terms we find the corresponding anomalous amplitude,
\begin{equation}
\hbar\Omega A_{tot}(t,E) = \frac{i\hbar}{2} \left\{ \frac{\partial S_{1} }{ \partial t } \frac{\partial S_{2} }{ \partial E } - \frac{\partial S_{1} }{ \partial E } \frac{\partial S_{2} }{ \partial t } \right\}\,,
\label{21}
\end{equation} 
\noindent
satisfying Eq.\,(\ref{05}) with $S$ is replaced by $S_{tot} = S_{1}S_{2}$.
Substituting the adiabatic expansion for $S_{in,tot}$ into Eq.\,(\ref{01}) we find the noise power, ${\cal P}_{l,tot}(\omega)$, generated by the driven double-cavity capacitor for $k_{B}\theta \gg \hbar\Omega$ or $|\omega| \gg \Omega$, given by Eq.\,(\ref{11}), where we replace $\{\nu^2(E)\}_{l}$ with $\left\{\left[ \nu_{1}(E) + \nu_{2}(E) \right]^{2} \right\}_{l}$. 
Then comparing it with Eq.\,(\ref{19}) we conclude that there is no a simple relation between $I_{E,tot}$ and ${\cal P}_{l,tot}(\omega)$ similar to Eq.\,(\ref{13}) for a single-cavity. 
\\ \indent
In the quantized emission regime, $\Gamma_{j}\Omega \ll 1$, the quantity ${\cal P}_{l,tot}(\omega)$ is roughly the sum of the noise powers produced by each cavity separately, if they emit particles at different times, $\left|t_{1\alpha} - t_{2\beta}\right| \gg \Gamma_{1} + \Gamma_{2}$.
Whenever the two particles are emitted simultaneously, the noise is enhanced no matter whether these particles are of the same kind (two electrons or two holes) or whether they are different 
(an electron-hole pair). 
This is in striking contrast to the heat current, $I_{E,tot}$, which is enhanced in the former case and is suppressed in the later case. 
In Fig.\,\ref{fig2} we show both the noise and the heat current produced by the double-cavity capacitor in the quantized emission regime, when each cavity emits one electron and one hole 
during the period. 
If we change the phase lag, $\delta = \delta_{2} - \delta_{1}$, between $U_{1}(t)$ and $U_{2}(t)$
then the relative time when cavities emit particles changes. 
The left peak corresponds to a two-particle emission, while the right peak and dip correspond to an electron-hole pair emission.
\\ \indent
\textit{Conclusion.}-- We have explored energetics and correlation properties of a dynamical quantum capacitor functioning as a single- or two-particle emitter. 
We showed that at low temperatures the relaxation resistance quantum $R_{q}$ defines the heat production in both linear and quantized emission regimes.
This allows to estimate heat flow from purely electrical measurements. 
We found that the pair of electrons emitted by the double-cavity capacitor carries an energy larger than that of two separately emitted electrons. 
This is a general effect inherent to multi-particle emitters. 
\\ \indent
We thank J. Splettstoesser for useful discussion.
We acknowledge the support of the Swiss NSF, the program for MANEP, and the EU project SUBTLE.

\end{document}